\documentclass[twocolumn,showpacs]{revtex4}

\usepackage{graphicx}
\usepackage{dcolumn}
\usepackage{bm}

\begin{document}


\title{Multiple Scattering in a Vacuum Barrier from Real-Space
Wavefunctions
       }

\author{Werner A. Hofer}
\email{w.hofer@liverpool.ac.uk} \affiliation{
         Donostia International Physics Center, San Sebastian,
         Spain, and \\
         Surface Science Research Centre,
         University of Liverpool, Liverpool L69 3BX, Britain}

\begin{abstract}
We have developed a method to simulate multiple electron
scattering in a vacuum barrier using real-space single-electron
wavefunctions of the separate surfaces. The tunnelling current is
calculated to first order in the Dyson series. In zero order, we
find a result which differs from the usual Bardeen approach by the
decay constants of the wavefunctions, entering the description as
individual weights of tunnelling transitions. To first order we
find multiple electron scattering, which can be formulated in
terms of Bardeen matrices. Here, we also derive a first-principles
formulation for the interaction energy between the two surfaces.
With this method the tunnelling current can in principle be
computed  to any order in the Dyson expansion.
\end{abstract}

\maketitle

From a theoretical point of view a tunnelling
electron, e.g. in a scanning tunnelling microscopy
measurement, is part of a system comprising two
infinite metal leads and an interface, consisting
of a vacuum barrier and, optionally, a molecule or
a cluster of atoms with different properties than
the infinite leads. The system can be said to be
open - the number of charge carriers is not
constant - and out of equilibrium - the applied
potential and charge transport itself introduce
polarizations and excitations within the system.
The theoretical description of such a system has
advanced significantly over the last years, to date
the most comprehensive description is based either
on a self-consistent solution of the
Lippman-Schwinger equation \cite{lang02} or on the
non-equilibrium Green's function approach
\cite{meir92,datta95,taylor01,brandbyge02,garcia03}.
Inelastic effects within e.g. a molecule-surface
interface can be included by considering multiple
electron paths from the vacuum into the surface
substrate \cite{lorente00}. Within the vacuum
barrier itself, inelastic effects play an
insignificant role. Here, as in most experiments in
scanning tunneling microscopy, the problem can be
reduced to the description of the tunnelling
current between two leads - the surface $S$ and the
tip $T$ - thought to be in thermal equilibrium. The
bias potential of the circuit is in this case
described by a modification of the chemical
potentials of surface and tip system, symbolized by
$\mu_S$ and $\mu_T$. This reduces the tunnelling
problem to the Landauer-B\"uttiker formulation
\cite{landauer85,datta95}, or :
\begin{eqnarray}\label{lb01}
I &\propto&  \int_{-\infty}^{+\infty} dE \left[f(\mu_S,E) -
f(\mu_T,E)\right] \times \nonumber \\
&\times& Tr \left[\Gamma_T(E) G^R(E) \Gamma_S(E) G^A(E)\right]
\end{eqnarray}

$f$ denotes the Fermi distribution functions, $G^{R(A)}(E)$ and
$G^A(E)$ are the retarded (advanced) Greens functions of the
barrier, the $\Gamma_{S(T)}$ are the surface and tip contacts.
They correspond to the difference of retarded and advanced self
energy terms of surface and tip; we define them by their relation
to the spectral function $A_{S(T)}$ of the surface (tip)
\cite{datta95}:
\begin{equation}\label{sf01}
A_{S(T)} = i \left[G_{S(T)}^R - G_{S(T)}^A\right] = G_{S(T)}^R
\Gamma_{S(T)} G_{S(T)}^A
\end{equation}

Here, the explicit energy dependency has been
omitted for brevity. At present, these equations
are evaluated within localized basis sets, and in a
matrix representation. From a theoretical point of
view this requires to either represent the
electronic properties of the two surfaces also in a
localized representation
\cite{brandbyge02,garcia03}, or to transform the
plane wave basis set of most density functional
methods to a local basis. The use of local basis
sets compromises the numerical accuracy in the
tunnelling barrier, since the vacuum tails of the
surface wavefunctions decay too rapidly: the
constant current contours in this case are too
close to the surface.

In this Letter we present a formulation of the problem which is
based on the Green's functions of the two surfaces, given in a
real space representation based on the electronic eigenstates of
the two systems. We show how the multiple scattering formalism
described in Eq. \ref{lb01} can be evaluated in real space, and
how it relates to the perturbation expansion of the tunnelling
problem. We start with an eigenvector expansion of the two surface
Green's functions, given by:
\begin{equation}\label{gf01}
G_{S}^{R(A)}({\bf r}_1,{\bf r}_2,E) = \sum_i \frac{\psi_i({\bf
r}_1) \psi_i^*({\bf r}_2)}{E - E_i +(-) i \eta}
\end{equation}
\begin{equation}\label{gf02}
G_{T}^{R(A)}({\bf r}_1,{\bf r}_2,E) = \sum_j
\frac{\chi_j({\bf r}_1) \chi_j^*({\bf r}_2)}{E -
E_j +(-) i \epsilon}
\end{equation}

Throughout this paper the wavefunctions $\psi$ and
$\chi$ denote the Kohn-Sham states of surface and
tip, respectively, resulting from a density
functional calculation. The spectral function $A_S$
describes the charge density matrix, from Eq.
\ref{sf01} we find:
\begin{equation}\label{sp01}
A_S ({\bf r}_1,{\bf r}_2,E) = 2 \eta \sum_i
\frac{\psi_i({\bf r}_1) \psi_i^*({\bf r}_2)}{(E -
E_i)^2 + \eta^2}
\end{equation}

The spectral function is related to $\Gamma_S$ by Eq. \ref{sf01}.
With the ansatz for $\Gamma_S$:
\begin{equation}
\Gamma_S ({\bf r}_3,{\bf r}_4,E) = C \sum_j \psi_j({\bf r}_3)
\psi_j^*({\bf r}_4)
\end{equation}
where $C$ is a constant, we perform the double volume integration
of Eq. \ref{sf01}. In this case the orthogonality of surface
states reduces the expression to a compact form:
\begin{eqnarray}\label{ga01}
&C& \int d r_3 d r_4 G_S^R ({\bf r}_1,{\bf r}_3,E)
\Gamma_S ({\bf r}_3,{\bf r}_4,E) G_S^A ({\bf
r}_4,{\bf r}_2,E) \nonumber \\&=& C \sum_{ijk}
\frac{\psi_i({\bf r}_1) \psi_k^*({\bf r}_2)
\delta_{ij} \delta_{jk}}{(E - E_i + i \eta)(E - E_k
- i\eta)}
\end{eqnarray}
Comparing the result with Eq. \ref{sp01} we find
for the contacts of surface and tip:
\begin{eqnarray}\label{sf10}
\Gamma_S = 2 \eta \sum_k \psi_k({\bf r}_3) \psi_k^*({\bf r}_4)
\quad \Gamma_T = 2 \epsilon \sum_i \chi_i({\bf r}_1) \chi_i^*({\bf
r}_2)
\end{eqnarray}

For the construction of the Greens function in the
barrier we use the fact that the charge density is
known from the separate calculation of surface and
tip. In the limit of weak coupling, the total
charge density of the interface is given by:
\begin{eqnarray}
n({\bf r}_1,E) &=& \sum_i \psi_i({\bf r}_1) \psi_i^*({\bf r}_1)
\delta (E-E_i) + \nonumber \\ &+& \sum_j \chi_j({\bf r}_1)
\chi_j^*({\bf r}_1) \delta(E-E_j)
\end{eqnarray}

The surface Green's function fulfills the boundary conditions at
the interface between surface and vacuum. The tip Green's function
at this position can be considered as a weak perturbation, which
does not alter the electronic properties of the surface.
Conversely, the same statement holds for the interface between the
vacuum and the tip with respect to the tip Green's function. This
allows to construct the zero order approximation for the Green's
function of the vacuum barrier as the sum of surface and tip
Green's functions, or:
\begin{equation}
G_{(0)}^{R(A)}({\bf r}_1,{\bf r}_2,E) =
G_S^{R(A)}({\bf r}_1,{\bf r}_2,E) + G_T^{R(A)}({\bf
r}_1,{\bf r}_2,E)
\end{equation}

Now all the necessary components for calculating the trace in the
non equilibrium formalism are given in terms of the real space
surface and tip wavefunctions. The trace involves the following
integral:
\begin{eqnarray}
&Tr& [\Gamma_T G^R \Gamma_S G^A] = \\&=& \int d r_{1-4}
\Gamma_T({\bf r}_1,{\bf r}_2) G_{(0)}^R({\bf r}_2,{\bf r}_3)
\Gamma_S({\bf r}_3,{\bf r}_4) G_{(0)}^A({\bf r}_4,{\bf r}_1)
\nonumber
\end{eqnarray}

The integrations have to be performed successively.
We start with the first integral, which involves:
\begin{eqnarray}
&2\eta& \int d r_4 \sum_m \psi_m({\bf r}_3) \psi_m^*({\bf r}_4)
\times
\\
&\times& \left[ \sum_n\frac{\psi_n({\bf r}_4)\psi_n^*({\bf
r}_1)}{E - E_n - i\eta} + \sum_n\frac{\chi_n({\bf
r}_4)\chi_n^*({\bf r}_1)}{E - E_n - i\eta}\right] \nonumber
\end{eqnarray}

Here, we use the physical condition of weak coupling to simplify
the evaluation. In this limit the overlap integrals between states
$\psi_m$ and $\chi_n$ will be substantially smaller than overlaps
between $\psi_m$ and $\psi_n$. Then the integral is reduced due to
the orthogonality of the wavefunctions to the simple form:
\begin{equation}
\int d r_4 = 2\eta \sum_m \frac{\psi_m({\bf r}_3)\psi_m^*({\bf
r}_1)}{E - E_m - i\eta}
\end{equation}

The same condition leads to the compact result for the second
integral:
\begin{equation}
\int d r_3 = 2\eta \sum_k \frac{\psi_k({\bf r}_2)\psi_k^*({\bf
r}_1)}{(E - E_k)^2 + \eta^2}
\end{equation}

The last two integrals involve the overlap between surface and tip
states. The final relation for the trace reads then:
\begin{equation}
Tr = \sum_{ik}\frac{4 \eta \epsilon}{(E - E_k)^2 +
\eta^2}\underbrace{\left|\int d r_1 \chi_i^*({\bf r}_1)\psi_k({\bf
r}_1)\right|^2}_{=: |A_{ik}|^2}
\end{equation}

The calculation of the matrix elements involves an integration
over infinite space, which cannot directly be performed. To
convert the volume integrals into surface integrals we use the
fact that the vacuum states of surface and tip are free electron
solutions with characteristic decay constants, complying with the
vacuum Schr\"odinger equation.
\begin{eqnarray}\label{se01}
\frac{\hbar^2}{2 m}\left(\nabla^2 - \kappa_i^2\right) \chi_i({\bf
r}) = 0 \Rightarrow \chi_i({\bf r}) = \frac{\nabla^2}{\kappa_i^2}
\chi_i({\bf r}) \\
\frac{\hbar^2}{2 m}\left(\nabla^2 - \kappa_k^2\right) \psi_k({\bf
r}) = 0 \Rightarrow \psi_k({\bf r}) = \frac{\nabla^2}{\kappa_k^2}
\psi_k({\bf r})
\end{eqnarray}

In addition we make use of the following identities:
\begin{eqnarray}
\chi_i^* \nabla^2 \psi_k = \nabla (\chi_i^* \nabla \psi_k) -
\nabla \chi_i^* \nabla \psi_k \nonumber \\
\psi_k \nabla^2 \chi_i^* = \nabla (\psi_k \nabla \chi_i^*) -
\nabla \chi_i^* \nabla \psi_k \nonumber
\end{eqnarray}

After some trivial manipulations, and making use of Gauss'
theorem, this leads to the result:
\begin{equation}
A_{ik} = \frac{1}{\kappa_k^2 - \kappa_i^2}\int dS
\left[\chi_i^*({\bf r})\nabla \psi_k({\bf r}) - \psi_k({\bf r})
\nabla \chi_i^*({\bf r})\right]
\end{equation}

The surface integral is well known; apart from the universal
constant $\hbar^2/2m$ it describes the tunneling matrix element in
the perturbation approach \cite{bardeen61,hofer03a}. Denoting the
surface integral by $M_{ik}$, we obtain for the zero order
expansion of the Green's functions the trace:
\begin{eqnarray}
Tr = 4 \sum_{ik}\frac{\eta \epsilon}{(E - E_k)^2 +
\eta^2}\left|\frac{M_{ik}}{\kappa_k^2 - \kappa_i^2}\right|^2
\end{eqnarray}

Considering that the states will have discrete energy levels $E =
E_k$ and setting $\eta \approx \epsilon$ the limit $\eta, \epsilon
\rightarrow 0^+$ is easy to evaluate. To check for consistency, we
also calculated the trace if both contacts are part of the
surface, described by $\Gamma_S$ and the Green's function $G_{S}$.
The calculation yields:
\begin{equation}
Tr = 4 \sum_k \frac{\eta \epsilon}{(E - E_k)^2 + \eta^2}
\end{equation}

Considering, that the transition probability in this case must be
unity, we conclude that the trace has to be scaled by a factor of
$1/4$ to yield the transition probability.  The final formulation
for the tunnelling current thus reads:
\begin{equation}
I_{(0)} = \frac{2 e}{h}
\sum_{ik}\int_{-\infty}^{+\infty}dE
\left[f(\mu_S,E) - f(\mu_T,E)\right]\left|\frac{
M_{ik}}{\kappa_k^2 - \kappa_i^2}\right|^2
\end{equation}

Note that $|M_{ik}/(\kappa_k^2 - \kappa_i^2)|^2$ is a
dimensionless value. It describes, like in the
Landauer-B\"utticker equation \cite{landauer85}, the transition
probability $T_{ik} = t_{ik}^+ t_{ik}$. Even though the
formulation looks quite similar to the standard results in
perturbation theory, it contains some decisive differences. While
in the perturbation approach the condition of resonant tunnelling
is explicitly included by a delta functional, this is not the case
in the scattering formulation. Instead, the matrix elements are
divided by the difference between the decay constants of the two
wavefunctions.  The decay constants describe the longitudinal
component of the electron momentum, they enter the description,
since they account for energy conservation via the vacuum
Schr\"odinger equations (see Eq. \ref{se01}). Their difference is
also related to the speed of charge transfer from one side of the
junction to the other. Given that the denominator differs for
every transition, this formulation describes, for the first time,
how individual contributions to the current have to be treated in
a proper manner.

We may compare this new formulation for the zero order tunnelling
current in the non equilibrium approach to the traditional
formulation based on Fermi's golden rule and the Bardeen method.
There we find for the current \cite{bardeen61,hofer03a}:
\begin{eqnarray}
I_{b} = \frac{4 \pi e}{\hbar} \sum_{ik}\left[f(\mu_S,E_k) -
f(\mu_T,E_i)\right] \times \\ \times \left|- \frac{\hbar^2}{2m}
M_{ik}\right|^2 \delta(E_i - E_k)\nonumber
\end{eqnarray}

Neglecting initially the decay constants and using
atomic units ($e = m = \hbar = 1$) the scattering
approach yields currents for $E_i = E_k$ which are
only about one tenth the currents obtained in the
approach based on Fermi's golden rule. However,
this is not the full picture. The decay constants
for metals are in the order of 0.5 to 0.6 atomic
units. Depending on the difference in work
functions of surface and tip, the current in the
scattering formulation will be increased by the
denominator. The increase will be moderate for e.g.
tungsten tips and noble metal surfaces, where the
workfunctions differ by about 1 to 2 eV. But even
in this case it should more than compensate for the
initial difference. The reason seems to be that the
time of transition is not infinite, as assumed in
the derivation of Fermi's golden rule, but finite.

The approach can be extended to higher orders. In the first order
expansion of the Dyson series the Green's function is given by:
\begin{equation}
G_{(1)}^R = G_{(0)}^R + G_{(0)}^R V G_{(0)}^R
\end{equation}

To calculate the first order Green's function for systems out of
equilibrium, the equation has to be solved self-consistently
\cite{lang02,brandbyge02}. Under tunnelling conditions, however,
the leads are in thermal equilibrium and the systems only weakly
coupled. $V$ in this case is the tip potential $V_T$ within the
vacuum barrier \cite{chen93}. Self consistency can in principle
also be achieved by basing the calculation on the Kohn-Sham states
$\psi$ and $\chi$ of charged surfaces \cite{alavi03}.
\begin{eqnarray}\label{gf10}
G_{(1)}^R({\bf r}_1,{\bf r}_2) = \nonumber \\
G_{(0)}^R({\bf r}_1,{\bf r}_2) + \int d r_{3}
G_{(0)}^R({\bf r}_1,{\bf r}_3) V_T({\bf r}_3)
G_{(0)}^R({\bf r}_3,{\bf r}_2)
\end{eqnarray}

Setting $\epsilon \approx \eta$ and with the
shortcut $f_{ik}^{\pm} = (E - E_i \pm i\eta)(E -
E_k \pm i\eta)$ the first order Green's function
depends on three additional terms:
\begin{eqnarray}
&+& \sum_{ik}\frac{\psi_i({\bf r}_1)\psi_k^*({\bf r}_2)}{f_{ik}^+}
\int d r_3 \psi_i^*({\bf
r}_3)V_T({\bf r}_3)\psi_k({\bf r}_3) \nonumber \\
&+& \sum_{ik}\frac{\psi_i({\bf r}_1)\chi_k^*({\bf r}_2)}{f_{ik}^+}
\int d r_3 \psi_i^*({\bf
r}_3)V_T({\bf r}_3)\chi_k({\bf r}_3)  \\
&+& \sum_{ik}\frac{\chi_i({\bf r}_1)\psi_k^*({\bf
r}_2)}{f_{ik}^+} \int d r_3 \chi_i^*({\bf
r}_3)V_T({\bf r}_3)\psi_k({\bf r}_3) \nonumber
\end{eqnarray}

On the one hand the three additional contributions describe
excitations of surface electrons. These contributions can be
included in the formulation by a suitable adaptation of many-body
theory. On the other hand they describe the scattering across the
barrier due to the vicinity of surface and tip. These matrix
elements, however, are well known. Apart from an insignificant
contribution within the surface system, they are just the Bardeen
matrix elements in the zero order expansion, or
\cite{chen93,hofer03a}:
\begin{equation}
\int d r_3 \chi_i^*({\bf r}_3)V_T({\bf
r}_3)\psi_k({\bf r}_3) = - \frac{\hbar^2}{2 m}
M_{ik}
\end{equation}

Limiting the evaluation in the following to scattering events
across the barrier, we may write for the first order Green's
function the expansion:
\begin{eqnarray}\label{gf11}
G_{(1)}^R &=& G_{(0)}^R - \\
&-&
\frac{\hbar^2}{2m}\sum_{ik}\left[\frac{\chi_i({\bf
r}_1)M_{ik}\psi_k^*({\bf r}_2)}{f_{ik}^+} +
\frac{\psi_i({\bf r}_1)M_{ki}^*\chi_k^*({\bf
r}_2)}{f_{ik}^+}\right] \nonumber
\end{eqnarray}

The trace now has to be evaluated for the additional
contributions. The same strategy as used for the zero order
expansion leads in the limit of $\eta \rightarrow 0^+$ to only one
surviving multiple scattering term, given by:
\begin{eqnarray}
\left(\frac{\hbar^2}{2 m}\right)^2 \sum_{iklm}
\frac{M_{ik}M_{lk}^*M_{ml}M_{im}^*}{(\kappa_l^2 -
\kappa_k^2)(\kappa_i^2 - \kappa_m^2)(E_i - E_k)(E_m - E_l)}
\nonumber
\end{eqnarray}

The current through the tunnelling barrier in the first order
expansion of the scattering series is therefore:
\begin{eqnarray}
I_{(1)} = I_{(0)} + \frac{2 e}{h}\sum_{iklm} \int
dE \left[f(\mu_S,E) -
f(\mu_T,E)\right] \times \,\\
\times  \left(\frac{\hbar^2}{2m}\right)^2
\frac{M_{ik}}{(E_i - E_k)}\frac{
M_{lk}^*}{(\kappa_l^2 -
\kappa_k^2)}\frac{M_{ml}}{(E_m -
E_l)}\frac{M_{im}^*}{(\kappa_i^2 -
\kappa_m^2)}\nonumber
\end{eqnarray}

Note that the product in the bottom line is again dimensionless
and describes the transition probability in the multiple
scattering case. The expression contains some interesting
features. Firstly, the dynamics of electron scattering are fully
included by the decay constants in the denominator; they describe
the difference in propagation velocity for the two different
directions. It is also quite interesting, that the expression
reaches a maximum, if the decay constants are roughly equal. This
indicates a potential resonance in tunnelling circuits, which
should be detectable by accurate comparisons between experiment
and theory. Secondly, the denominator contains the energy
differences to the intermediate states in the multiple scattering
process. This is quite understandable from perturbation theory,
where e.g. the interaction energies scale with the inverse of the
energy difference. In this case the intermediate states have to be
evaluated over the whole energy range.

Finally, we consider the interaction energy between the surface
and the tip in the low coupling limit. It has been shown recently
by an analysis of first order perturbation expressions for the
tunnelling current and the interaction energy, that the two
variables should be linear with each other. From the first order
Green's function we may construct the density matrix $\hat{n}$.
The interaction energy is then \cite{hofer03b}:
\begin{equation}
E_{int} = Tr [\hat{n} V] = \frac{i}{2 \pi} Tr
\left[\left(G_{(1)}^R - G_{(1)}^A\right) V\right]
\end{equation}

With $G^R_{(1)}$ and  $G^A_{(1)}$ from Eq. \ref{gf11} this leads
to the following result:
\begin{equation}
E_{int} = \left(\frac{\hbar^2}{2m}\right)^2
\sum_{ik} \int \frac{dE}{2 \pi}
\left(\frac{i}{f_{ik}^+} -
\frac{i}{f_{ik}^-}\right) \left( |M_{ik}|^2 +
|M_{ki}|^2\right)
\end{equation}

The energy integration in this case is not trivial, since a term
containing $1/f_{ik}^{\pm}$ will be singular for $\eta \rightarrow
0^+$ and $E = E_i$ or $E = E_k$. This is due to the missing
resistance of electrons upon reflection at the surfaces. If a
resistance term $\eta$ is introduced in the same manner as e.g.
for the contacts $\Gamma$ (see Eq. \ref{sf10}), then the result
converges, and we obtain:
\begin{equation}
E_{int} = - \frac{1}{\pi}
\left(\frac{\hbar^2}{2m}\right)^2 \sum_{ik}
\frac{|M_{ik}|^2 + |M_{ki}|^2}{|E_i - E_k|}
\end{equation}

Physically, we may relate this addition to the fact that
transiting electrons will lead to charge oscillations in the
surface layers, depending on the conductance properties of the
leads. The calculation of the interaction energy only involves the
computation of the tunnelling matrix elements. And, as shown
previously, the interaction energy will therefore be proportional
to the tunnelling current \cite{hofer03b}.

To summarize, tunnelling currents and interaction energies can be
calculated in real space within the non equilibrium Green's
function formalism based on the separate wavefunctions of surface
and tip. The zero order expansion is similar to the traditional
Bardeen approach, even though we find that each transition carries
an individual weight. This also implies that resonances between
surface and tip will play an essential role under specific
material conditions. The first order expansion describes multiple
scattering across the tunnelling barrier. In this case we can
derive a first principles formulation for the interaction energy
between the two surfaces.

The author acknowledges helpful discussions with A. Arnau, P.
Echenique,  M. Johnson, N. Lorente, J. Soler, and L. Wirtz. He
also thanks the Royal Society for the award of a University
Research Fellowship.

\end{document}